\documentclass{JHEP3}
\usepackage{amssymb,amsfonts,amsmath,amsthm,latexsym,verbatim}
\newcommand{\be}{\begin{equation}}
\newcommand{\ee}{\end{equation}}
\newcommand{\bea}{\begin{eqnarray}}
\newcommand{\eea}{\end{eqnarray}}
\newcommand{\Mc}{{\cal M}_c}
\newcommand{\Mct}{\Mc(\tau)}
\newcommand{\al}{\alpha}
\newcommand{\zm}{z_{m}}
\newcommand{\zn}{z_{n}}

\newcommand{\bzn}{\bar z_{n}}
\newcommand{\bzl}{\bar z_{l}}
\newcommand{\pd}{\partial}
\newcommand{\hs}{h_{ijk}}
\newcommand{\ta}{\theta}
\newcommand{\xp}{x^+}

\title{\center{On metric geometry of conformal moduli spaces of four-dimensional superconformal theories}}

\author{
    Vadim Asnin\footnotemark[1]\\
    Racah Institute of Physics\\
    Hebrew University \\
    Jerusalem 91904,
    Israel\\

    \footnotemark[1] {\tt vadim.asnin@mail.huji.ac.il\\}

        }

\abstract{Conformal moduli spaces of four-dimensional superconformal
theories obtained by deformations of a superpotential are
considered. These spaces possess a natural metric (a Zamolodchikov
metric). This metric is shown to be Kahler. The proof is based on
superconformal Ward identities.}
\begin{document}
\section{Introduction}
Since the seminal work of Leigh and Srassler \cite{Leigh:1995ep} who
showed that four-dimensional superconformal theories often come in
continuous families called conformal moduli spaces there has been a
considerable progress in a study of these spaces.

An important progress has been made in connection with the $AdS/CFT$
correspondence. Supergravity duals of superconformal deformations
were studied perturbatively in \cite{Aharony:2002hx} and more
generally in \cite{Kol:2002zt}, where it was shown that a dimension
of the conformal moduli space in the supergravity case is equal to a
certain well-defined index. Some examples of exact supergravity
duals of superconformal theories became accessible with a specific
construction of families of Sasaki-Einstein spaces. In
\cite{Lunin:2005jy} there was found an exact dual of a so called
$\beta$-deformation of the ${\cal N}=4$ theory, which is a
particular kind of a general Leigh-Strassler deformation. It was
shown also that all toric quiver gauge theories admit
$\beta$-deformations \cite{Benvenuti:2005wi} which have supergravity
duals \cite{Butti:2007aq}.

Any conformal moduli space has a natural metric defined on it,
namely, the Zamolodchikov metric \cite{Zamolodchikov:1986gt}. It is
defined in terms of exactly marginal operators $O_i(x)$. This metric
turns out to be Kahler for the exact supergravity dual of a
$\beta$-deformation \cite{Lunin:2005jy} and was shown to be Kahler
in general in the supergravity \cite{Tachikawa:2005tq}. All these
results concerned $SU(N)$ field theories with infinite $N$ and it
was not clear whether they hold for $1/N$ corrections as well.

On the other side, there are well established results about
two-dimensional conformal theories with $(2,2)$ supersymmetry, which
are analogs of four-dimensional ${\cal N}=1$ superconformal
theories. The conformal moduli spaces of these theories are coset
spaces and the metric on them is Kahler \cite{Seiberg:1988pf}.

In this paper we show that the Zamolodchikov metric on conformal
moduli spaces of four-dimensional superconformal theories with 8
supercharges is Kahler (as it was initially conjectured in
\cite{AsninKol}). We use purely field theoretic methods, so the
results hold for any $N$. The main tool is the superconformal Ward
identities which constrain a possible form of various correlators of
exactly marginal operators. As an example we work with a space of
deformations of ${\cal N}=4$ theory, but our results are general.

The proof of a Kahlerity is given in section \ref{KahleritySection}.
Section \ref{Discussion} is a brief summary and a discussion of open
questions.


\section{Proof of the Kahlerity of the Zamolodchikov
metric}\label{KahleritySection} We consider a 4D ${\cal{N}}=1$
superconformal theory obtained by a Leigh-Strassler
\cite{Leigh:1995ep} deformation of the superpotential of the
${\cal{N}}=4$ theory, although our treatment is applicable to other
theories as well. The Lagrangian of the theory is \begin{multline}
L=\sum\limits_{i=1}^3 Tr\int d^4\ta\, e^{-V}\bar\Phi_i
e^V\Phi_i+\frac{\tau}{32\pi i}\,Tr\int d^2\ta\, W^2+\\+Tr\int
d^2\ta\,\bigl(h_0\epsilon^{ijk}\Phi_i\Phi_j\Phi_k+h_1
f^{ijk}\Phi_i\Phi_j\Phi_k+h_2(\Phi_1^3+\Phi_2^3+\Phi_2^3)\bigr)+c.c,
\end{multline} where $\tau$ is a usual combination of a coupling constant and a $\theta$
angle: $\tau=\frac{\theta}{2\pi}+\frac{4\pi i}{g^2}$ and
$\epsilon^{ijk}$ and $f^{ijk}$ are antisymmetric and symmetric
$SU(3)$ invariants respectively. The coupling constants $h_{0,1,2}$
are complex-valued, however their common phase is of no importance
since it can be compensated by a redefinition of superfields.

The argument of Leigh and Strassler shows that this theory is
conformal as long as the coupling constants satisfy the equation \be
\gamma(\tau, h_0, h_1,
h_2,\bar{\tau},\bar{h}_0,\bar{h}_1,\bar{h}_2)=0,\label{LSequation}\ee
where $\gamma$ is an anomalous dimension of chiral superfields (the
form of the Lagrangian guarantees that all three chiral superfields
have he same dimension because of various discrete symmetries
between the three superfields). Solutions of this equation which
differ by a common phase of $h_{0,1,2}$ should be identified. The
$U(1)$ quotient space obtained this way is a conformal moduli space
$\Mc$.\footnote{There are additional discrete identifications of
points on $\Mc$ due to the $SL(2,\mathbb{Z})$ duality.} It has a
real dimension 6. In what follows we take the gauge coupling
constant $\tau$ to be fixed and allow the superpotential couplings
$h_k$ to change so that the eq. \ref{LSequation} would be satisfied,
so we concentrate on four-dimensional slices of $\Mc$, which we will
denote by $\Mc(\tau)$.

The first point that we need to establish is a fact that there are
(local) complex coordinates on $\Mct$ such that the coupling
constants $h_{0,1,2}$ are their holomorphic functions. To see this
we notice that the definition of $\Mct$ resembles a definition of a
complex projective space $\mathbb{CP}_n$ is terms of an affine space
$\mathbb{C}_{n+1}$. Indeed, in both cases the definition includes a
single real equation (\ref{LSequation} for $\Mct$ and an equation of
a sphere for $\mathbb{CP}_n$) and a $U(1)$ which rotates phases of
coordinates of the ambient affine complex space. We need to choose
coordinates on $\Mct$ which are invariant under this $U(1)$. We know
that the point $h_0=g,h_1=h_2=0$ belongs to $\Mct$. We choose some
neighborhood of this point and analogously to $\mathbb{CP}_n$ define
local complex coordinates $\zn$, $n=1,2$ on $\Mc(\tau)$ to be
$z_{1,2}=h_{1,2}/h_0$. In order to see that these are good
coordinates we need to show that having chosen $z$'s, we can
reconstruct $h$'s without any need in $\bar{h}$'s. Having chosen
$z$'s, we get $h_{1,2}=z_{1,2}\,h_0$, and the eq. \ref{LSequation}
becomes \be\gamma(\tau, \,h_0, \,z_1\,h_0,
\,z_2\,h_0,\,\bar{\tau},\,\bar{h}_0,\,\bar{z}_1\,\bar{h}_0,\,\bar{z}_2\,\bar{h}_0)=0.\ee
In this eq. a phase of $h_0$ is not relevant because it can be
changed by a $U(1)$ rotation. We can choose $h_0$ to be real, for
example,  and then the last eq. becomes an equation for $|h_0|$. The
eq. is satisfied for $z_{1,2}=0$ (the solution is $h_0=g$) and by
the implicit function theorem it has solutions for $z_{1,2}$ being
sufficiently close to 0 (the conditions of the theorem are satisfied
since for small coupling constants the expansion of $\gamma$ is
$\gamma\sim|h_0|^2+|h^1|^2+|h_2|^2-g^2$ with some coefficients which
are not relevant for the argument). So we conclude that the required
choice of coordinates is possible.

The space of conformal deformations of the superpotential of ${\cal
N}=4$ theory was studied more generally in \cite{Aharony:2002hx,
Kol:2002zt}, without imposing any discrete symmetries on the
Lagrangian. In this case the superpotential is of a general form
$\hs Tr(\Phi_i\Phi_j\Phi_k)$ with symmetric complex-valued
coefficients $\hs$. Instead of a single anomalous dimension $\gamma$
in this formulation there appears a Hermitian matrix $\gamma_{ij}$
of anomalous dimensions, and instead of a single Leigh-Strassler
equation \ref{LSequation} there are 8 equations corresponding to the
traceless part of $\gamma_{ij}$. The number of independent coupling
constants is 10 and there is a global $SU(3)$ which identifies
different solutions, so the real dimension of the space of
superpotential deformations is again 4. As discussed in
\cite{Kol:2002zt}, close to the origin $\Mc(\tau)\approx{\bf
10}_{\mathbb{C}}/SL(3,\mathbb{C})$. This manifold is a solution of
the $D$-term constraint of the global $SU(3)$. A solution of a
$D$-term constraint is a result of a division of the ambient space
$10_{\mathbb{C}}$ by a complexified group, which is
$SL(3,\mathbb{C})$. It is also argued in \cite{Kol:2002zt} that
$\Mc(\tau)$ is a complex manifold, in an agreement with the argument
above.

 As discussed in the introduction, any conformal moduli space
is endowed with a natural metric, the Zamolodchikov metric
\cite{Zamolodchikov:1986gt}, defined in terms of correlators of
exactly marginal operators $O_i(x)$. In a $d$-dimensional theory a
mass dimension of exactly marginal operators is $d$ and a conformal
invariance fixes a form of a correlator of two such operators to be
$<O_i(x)O_i(y)>\sim|x-y|^{-2d}$. The correlator is similar to an
inner product of two tangent vectors, and the Zamolodchikov metric
is defined as a coefficient in this identity:\be
<O_i(x)O_i(y)>=g_{ij}|x-y|^{-2d}.\ee

We study the Zamolodchikov metric on the conformal moduli space of
4-dimensional superconformal theories $\Mct$. In this case the
supersymmetry imposes additional constraints on the metric. We will
show that these additional constraints make the metric Kahler, as
was proposed in \cite{AsninKol}.

As a starting point we use the fact that according to the
representation theory of superconformal groups \cite{Dobrev:1985qv}
there is a relation between a mass dimension and an $R$-charge of a
chiral primary field, namely $d=\frac32 r$ (a corresponding relation
for an antichiral primary field is $d=-\frac32 r$). A chiral field
is a lowest component of a chiral superfield, and corresponding
relations for highest ($F$-) components of chiral and antichiral
superfields are $d-1=\frac32(r+2)$ and $d-1=-\frac32(r-2)$
correspondingly. An exactly marginal operator is of dimension 4 and
has a vanishing $R$-charge, and these relations are satisfied.
Therefore any exactly marginal operator must be a linear combination
of $F$-components of chiral and antichiral superfields.

Next we notice that an exactly marginal operator is a derivative of
the Lagrangian w.r.t. a coordinate on $\Mc$. We have shown that
there are complex coordinates $\zn$ on $\Mct$ such that the coupling
constants $h_{k}$ are holomorphic functions of $\zn$. In these
coordinates the derivatives $\pd L/\pd\zn$ are proportional to
linear combinations of $\pd L/\pd h_{k}$. These derivatives are
$F$-components of chiral superfields (with no contribution of
antichiral ones). Correspondingly, $\pd L/\pd\bzn$ are
$F$-components of antichiral superfields.

We use now superconformal Ward identities to study restrictions that
supersymmetry imposes on the form of the Zamolodchikov metric. In
our coordinate system there are four different kinds of components
of the metric: $g_{m n}$, $g_{m\bar n}$, $g_{\bar mn}$, $g_{\bar
m\bar n}$. First of all we show that $g_{mn}=g_{\bar m\bar n}=0$. In
order to see this consider a correlator of two chiral superfields on
${\cal N}=1$ superspace. These superfields are functions of $\ta$
and a chiral coordinate $\xp_{\mu}=x_{\mu}+\frac
i2\ta\sigma\bar{\ta}$. Therefore the correlator is a function of
these coordinates:\be
<\Phi(\xp_1,\ta_1)\Phi(\xp_2,\ta_2)>=f(\xp_1,\xp_2,\ta_1,\ta_2).\label{2point}\ee
We are interested in the correlator of $F$-components, which is a
coefficient of $\ta^2\bar\ta^2$ on the RHS. In order to see that
this coefficient vanishes apply to both sides the superconformal
generator $S^{\dot\al}$ which in the chiral representation is (see
\cite{Gates:1983nr}) \be S^{\dot\al}\sim
\bar{\sigma}_{\mu}^{\al\dot{\al}}x^{+\mu}\pd_{\al}.\ee LHS of
\ref{2point} vanishes under the action of $S^{\dot{\al}}$, whereas
the coefficient of $\ta^2\bar\ta^2$ of RHS gets multiplied by
$\bar{\sigma}_{\mu}^{\al\dot{\al}}(x_1^{\mu}+x_2^{\mu})$. The
coefficient itself depends on $x_1$ and $x_2$ as $|x_1-x_2|^{-8}$
and therefore can vanish only if it is absent. So we see that the
correlator of the $F$-terms vanishes (actually, the only possible
term in a correlator of two chiral superfields is a contact term
\cite{Molotkov:1975gf, Aneva:1977rj, Osborn:1998qu}, we will discuss
this fact and use it later). This in turn means the term $g_{mn}$ of
the metric vanishes. Similar computation for antichiral superfields
shows that $g_{\bar m\bar n}$ vanishes as well. We see that the
metric in our coordinate system is Hermitian.

Next we show that the metric on $\Mct$ is Kahler. A definition of
the Kahler metric $g_{m\bar n}=\pd_m\pd_{\bar n}K$, where $K$ is a
Kahler potential, is equivalent to the integrability condition
$\pd_mg_{n\bar l}=\pd_ng_{m\bar l}$ (and its complex conjugate).
This is the condition that we are going to check.

The definition of the Zamolodchikov metric implies that
\begin{multline} \pd_mg_{n\bar l}=\pd_m<\frac{\pd L}{\pd\zn}(x)\frac{\pd
L}{\pd\bzl}(y)>|x-y|^8= \Biggl(<\frac{\pd L}{\pd\zn}(x)\frac{\pd
L}{\pd\bzl}(y)\int\,d^4u\,\frac{\pd L}{\pd\zm}(u)>\\+<\frac{\pd^2
L}{\pd\zm\pd\zn}(x)\frac{\pd L}{\pd\bzl}(y)>+<\frac{\pd
L}{\pd\zn}(x)\frac{\pd^2
L}{\pd\zm\pd\bzl}(y)>\Biggr)|x-y|^8\label{MainEq}\end{multline} A
few remarks about this equality are in order. First of all, the
second term on the RHS is symmetric under the interchange of $m$ and
$n$ and therefore satisfies the integrability condition. Next, as we
have shown, $\pd L/\pd\zn$ depends only on $z$'s and not on $\bar
z$'s, and therefore the last term on the RHS vanishes. Finally,
consider the first term which involves a three-point function of
exactly marginal operators. The conformal invariance in principle
allows an appearance of a term $|x-y|^{-4}|x-u|^{-4}|y-u|^{-4}$ in
the three-point function of operators of the mass dimension 4.
However, such a behavior is forbidden for exactly marginal operators
since it would lead to a non-trivial $\beta$-function of a
deformation operator in a perturbed theory. But there can appear
contact terms in the three-point function. Such contact terms  are
essential in a geometry of $\Mc$, as pointed out in
\cite{Kutasov:1988xb} for a two-dimensional case. We show now that
the superconformal invariance allows only for contact terms between
operators of the same chirality. Indeed, if there is a contact term
in the OPE of two $F$-terms of the same chirality:\be
F_m(x_1)F_n(x_2)\sim A_{mn}^k\delta^4(x_1-x_2)F_k(x_1)\ee then this
contact term can be promoted to a contact term on the superspace:
\be\Phi_m(x_1^+,\ta_1)\Phi_n(x_2^+,\ta_2)\sim
A_{mn}^k\delta^4(x_1^+-x_2^+)\delta^2(\ta_1-\ta_2)\Phi(x_1^+,\ta_1),\ee
with $\delta^2(\ta_1-\ta_2)\equiv(\ta_1-\ta_2)^2$. This is precisely
the contact term mentioned in our discussion of the two-point
functions. However, there is no supersymmetry covariant
generalization of a contact term between chiral and antichiral
superfields, and therefore a contact term between $F$-terms of
superfields of opposite chiralities is not consistent with the
supersymmetry. So indeed the only contact terms that are allowed are
those between fields of the same chirality.

This last point leads to the following observation about eq.
\ref{MainEq}. If the OPE between the exactly marginal operators is
$\pd_mL(x)\pd_nL(y)\sim A_{mn}^k\pd_kL(x)$ then the first term on
the RHS is $A_{mn}^k<\pd_kL(x)\pd_{\bar l}L(y)>$. Since
$A_{mn}^k=A_{nm}^k$ this term is symmetric under $m\leftrightarrow
n$.

We see that the whole RHS of eq. \ref{MainEq} is symmetric under
$m\leftrightarrow n$, and the integrability condition holds. We
conclude therefore that the Zamolodchikov metric on $\Mct$ is
Kahler.


\section{Discussion}\label{Discussion}

In this paper we considered a metric geometry of a conformal moduli
space of four-dimensional ${\cal N}=4$ theory which involves
deformations of a superpotential. We have shown that a natural
Zamolodchikov metric on this space is Kahler. The proof is based on
superconformal Ward identities for correlators of superfields on the
${\cal N}=1$ superspace. It involves three logical steps:
\begin{itemize}\item Show that there are complex coordinates on the
conformal moduli space such that (complex) coupling constants of the
theory depend on them holomorphically. This is not trivial because
$\Mct$ is a $U(1)$ quotient. But a similarity with a complex
projective space allows one to define holomorphic coordinates. \item
Show that the metric in these coordinates is Hermitian. The choice
of coordinates made in the previous step is crucial here and in the
next step because it guarantees that the exactly marginal operators
which span a tangent space to $\Mct$ are $F$-terms of superfields of
a definite chirality.\item Show that the integrability condition
which guarantees the existence of a Kahler potential is satisfied.
The main point here is a consideration of a three-point function of
exactly marginal operators. It is a combination of contact terms and
those of them that violate the integrability condition are forbidden
by the supersymmetry.\end{itemize}

There are a few remaining questions, though. One of them is to
derive explicitly the Kahler potential for the metric. Since the
metric is a two-point function of exactly marginal operators the
potential should be connected somehow to a partition function of the
theory, whose second derivatives w.r.t. coordinates of $\Mct$ give
correlators of exactly marginal operators integrated over the whole
space-time. However, the partition function of any supersymmetric
theory vanishes. The integrals of correlators diverge, but their
regulated versions vanish (for example, in a dimensional
regularization any power-like divergence is set to zero).  Therefore
a correct version of the Kahler potential is a partition function
regularized in some way which would break the supersymmetry. These
observations were made in \cite{AsninKol}. To derive the Kahler
potential is a subject of a future work.

Another issue to be studied is a generalization of our result to the
gauge coupling as well. The corresponding term in the Lagrangian is
again a chiral superfield and should be amenable to similar
treatment. The only possible problem is whether there is a choice of
``good" complex coordinate on $\Mc$ such that the gauge coupling
$\tau$ would be their holomorphic function. An intuition based on
the supergravity supports an affirmative answer.

\section*{Acknowledgements}
I thank B. Kol for participation at early stages of this work and O.
Aharony for discussions.

\end{document}